\documentstyle[12pt]{article}
\begin{document}
\input amssym.def
\input amssym
\hfuzz=5.0pt%

%
%
%
%
%
\def\vec#1{\mathchoice{\mbox{\boldmath$\displaystyle\bf#1$}}
{\mbox{\boldmath$\textstyle\bf#1$}}
{\mbox{\boldmath$\scriptstyle\bf#1$}}
{\mbox{\boldmath$\scriptscriptstyle\bf#1$}}}
\def\mbf#1{{\mathchoice {\hbox{$\rm\textstyle #1$}}
{\hbox{$\rm\textstyle #1$}} {\hbox{$\rm\scriptstyle #1$}}
{\hbox{$\rm\scriptscriptstyle #1$}}}}
\def\operatorname#1{{\mathchoice{\rm #1}{\rm #1}{\rm #1}{\rm #1}}}
\chardef\ii="10
\def\widehat{\mathaccent"0362 }
\def\widetilde{\mathaccent"0365 }
\def\vphi{\varphi}
\def\vrho{\varrho}
\def\vtheta{\vartheta}
\def\ih{{\i\over\hbar}}
\def\CD{{\cal D}}
\def\CL{{\cal L}}
\def\CP{{\cal P}}
\def\CV{{\cal V}}
\def\half{{1\over2}}
\def\bhalf{\hbox{$\half$}}
\def\viert{{1\over4}}
\def\bviert{\hbox{$\viert$}}
\def\oo{\operatorname{o}}
\def\OO{\operatorname{O}}
\def\SO{\operatorname{SO}}
\def\SU{\operatorname{SU}}
\def\ce{\operatorname{ce}}
\def\se{\operatorname{se}}
\def\Mc{\operatorname{Mc}}
\def\Ms{\operatorname{Ms}}
\def\ps{\operatorname{ps}}
\def\dfrac#1#2{\frac{\displaystyle #1}{\displaystyle #2}}
\def\pathint#1{\int\limits_{#1(t')=#1'}^{#1(t'')=#1''}\CD #1(t)}
\def\pathints#1{\int\limits_{#1(0)=#1'}^{#1(s'')=#1''}\CD #1(s)}
\def\hbarm{{\dfrac{\hbar^2}{2m}}}
\def\ints{\int_0^{s''}}
\def\intt{\int_{t'}^{t''}}
\def\limN{\lim_{N\to\infty}}
\def\dotv#1{\dot{\vec #1}^2}
\def\bbbr{{\rm I\!R}} 
\def\bbbn{{\rm I\!N}} 
\def\bbbz{{\mathchoice {\hbox{$\sf\textstyle Z\kern-0.4em Z$}}
{\hbox{$\sf\textstyle Z\kern-0.4em Z$}}
{\hbox{$\sf\scriptstyle Z\kern-0.3em Z$}}
{\hbox{$\sf\scriptscriptstyle Z\kern-0.2em Z$}}}}
\def\bbbone{{\mathchoice {\rm 1\mskip-4mu l} {\rm 1\mskip-4mu l}
{\rm 1\mskip-4.5mu l} {\rm 1\mskip-5mu l}}}
\def\d{\operatorname{d}}
\def\e{\operatorname{e}}
\def\i{\operatorname{i}}
 
\begin{titlepage}
\large
\thispagestyle{empty}
\centerline{\normalsize DESY 97 - 221 \hfill ISSN 0418 - 9833}
\centerline{\normalsize November 1997\hfill}
\centerline{\normalsize hep-th/9711119\hfill}
\message{TITLE:}
\message{ALTERNATIVE SOLUTION OF THE PATH INTEGRAL
          FOR THE RADIAL COULOMB PROBLEM}
\begin{center}
{\Large ALTERNATIVE SOLUTION OF THE PATH INTEGRAL
\vskip.05in
FOR THE RADIAL COULOMB PROBLEM}
\end{center}
\vskip.5in
\begin{center}
{\Large Christian Grosche}
\vskip.2in
{\normalsize\em II.\,Institut f\"ur Theoretische Physik}
\vskip.05in
{\normalsize\em Universit\"at Hamburg, Luruper Chaussee 149}
\vskip.05in
{\normalsize\em 22761 Hamburg, Germany}
\end{center}
\vfill
\normalsize
\bigskip\bigskip
\begin{center}
{ABSTRACT}
\end{center}
\smallskip
\noindent
{\small In this Letter I present an alternative solution of the
path integral for the radial Coulomb problem which is based on a
two-dimensional singular version of the Levi-Civita transformation.}
\bigskip\bigskip
\end{titlepage} 

 
\noindent
Ever since the development of the path integral by Feynman \cite{FEYb} it
has been challenge to calculate the propagator for the Coulomb potential.
Due to Feynman himself were the evaluation of the path integral for the
free particle and the harmonic oscillator, both belonging to a larger
class of path integrals which are called Gaussian Path Integrals because
they are based on the general quadratic Lagrangian. Nowadays more path 
integrals which go beyond the Gaussian form are known. They are the path 
integral solutions related to  the radial harmonic oscillator \cite{PI},
also called Besselian Path Integral, and the path integral solutions related 
to the P\"oschl-Teller and modified P\"oschl-Teller potential \cite{BJB}, also
called Legendrian Path Integrals. These Basic Path Integrals are the building 
blocks for a comprehensive Table of Path Integrals \cite{GRSc,GRSh}.

Since the path integral solutions of the one-dimensional and radial harmonic
oscillator were known for a long time, it was a challenge to evaluate the
path integral for the Coulomb potential, one of the most important physical
system in quantum mechanics. In 1979 Duru and Kleinert \cite{DKa,DKb} succeeded
in its calculation by means of a transformation known from astronomy
\cite{KUST}, and now called space-time transformation, ``Duru-Kleinert''
transformation, time substitution, etc.~\cite{CAST}--\cite{YODEWM}.
\footnote{Actually, the approach of Duru and Kleinert goes back to an idea by 
A.Barut presented in an informal seminal talk in Trieste 1978.
\par\noindent
The Kustaanheimo-Stiefel transformation is in fact a special case of the
Hurwitz transformation which works in the dimensions 1,2,4,8.}

In this contribution I would like to present an alternative evaluation of the
radial path integral of the Coulomb problem which is based on a singular
version of the Levi-Civita transformation, which can be seen as a 
two-dimensional version of the Kustaanheimo-Stiefel transformation \cite{KUST},
as it has been used for the two-dimensional Coulomb problem \cite{DKb,INOc}.
Let us consider the path integral for the Coulomb problem (C)  in radial
coordinates in three dimensions, where the propagator in $\bbbr^3$ has been 
separated according to
\begin{equation}
K^C(\vec x'',\vec x';T)=\sum_{l\in\bbbn}\sum_{n=-l}^lK^C_l(r'',r';T)
   Y^{n\,*}_l(\vtheta',\vphi') Y^{n\,*}_l(\vtheta'',\vphi'')\enspace,
\end{equation}
and $l$ is the usual angular momentum number, therefore
\begin{equation}
K_l^C(r'',r';T)=\dfrac1{r'r''}\pathint{r}\exp\Bigg\{\ih\intt\Bigg[
      \frac{m}2\dot r^2
      +\frac{e^2}r-\hbar^2\frac{l(l+1)}{2mr^2}\Bigg]dt\Bigg\}\enspace.
\label{KlC}
\end{equation}
Similarly as in \cite{DKa} I insert a factor ``one'' ($\epsilon=T/N$):
\begin{eqnarray}
  1&=&\sqrt{m\over2\pi\i\hbar T}\int_{\bbbr}dy''
     \exp\bigg(-\dfrac{m(y''-y')^2}{2\i\hbar T}\bigg)
  \nonumber\\   & &
  = \int_{\bbbr}dy''\limN\bigg(\dfrac{m}{2\pi\i\hbar\epsilon}\bigg)^{N/2}
    \prod_{j=1}^{N-1}\int_{\bbbr} dy_j\exp\Bigg(\dfrac{\i m}{2\epsilon\hbar}
    \sum_{j=1}^N(y_j-y_{j-1})^2\Bigg)\enspace.
\end{eqnarray}
We obtain for the path integral (\ref{KlC}) in the usual way
\begin{eqnarray} & &
K_l^C(r'',r';T)=\dfrac1{r'r''}\int_{\bbbr}dy''\pathint{r}\pathint{y}
   \qquad\qquad \qquad\qquad \qquad
   \nonumber\\   & &\qquad\qquad\times
   \exp\Bigg\{\ih\intt\Bigg[\frac{m}2(\dot r^2+\dot y^2)
      +\frac{e^2}r-\hbar^2\frac{l(l+1)}{2mr^2}\Bigg]dt\Bigg\}\enspace.
\label{KlCy}
\end{eqnarray}
I now consider in $\bbbr^2$ the mapping
\begin{equation}
 A(\vec u)=
 \left(\begin{array}{cc} u_1 &u_2\\ -u_2 &u_1\end{array}\right)\enspace,
 \qquad \vec u= \left(\begin{array}{c} u_1\\ u_2\end{array}\right)
 \in\bbbr^2\enspace.
\end{equation}
$A(\vec u)$ has the properties
\begin{equation}
  A(\vec u)\cdot \vec u= 
  \left(\begin{array}{c} u_1^2+u_2^2\\ 0\end{array}\right)\enspace,
  \qquad A^t(\vec u)\cdot A(\vec u)=(u_1^2+u_2^2)\bbbone
  =|\vec u|^2\bbbone\enspace.
\end{equation}
Therefore $A:\bbbr^2\mapsto\bbbr^+$. Hence $A$ is not one-to-one, and it is not
allowed to use the transformation $A$ in the path integral (\ref{KlCy})
in a na\"\ii ve way. Instead, following \cite{INOc}, we must define it on
midpoint coordinates in the following way ($i=1,2;j=1,\dots,N$):
\begin{equation}
\begin{array}{cc}
\Delta r_j&=2( \bar u_{1,j}\,\Delta u_{1,j}+\bar u_{2,j}\,\Delta u_{2,j})
\enspace,\\
\Delta\vec u_j&=2(-\bar u_{2,j}\,\Delta u_{1,j}+\bar u_{1,j}\,\Delta u_{2,j})
\enspace,\end{array} \qquad\begin{array}{cc}
\bar u_{i,j}  &=\bhalf(u_{i,j}+u_{i,j-1})\enspace,\\
\Delta u_{i,j}&=u_{i,j}-u_{i,j-1})       \enspace.\end{array} 
\label{Au}
\end{equation}
We then have $(\Delta r_j)^2+(\Delta y_j)^2=4(\bar{\vec u}_j)^2
(\Delta\vec u_j)^2$ and $A^t(\bar{\vec u}_j)\cdot A(\bar{\vec u}_j)= 
(\bar{\vec u}_j)^2\bbbone$. The Jocobian $J(\vec u)$ of the transformation 
$ A^t(\bar{\vec u}_j)$ for all $j=1,\dots,N$ reads $J(\bar{\vec u}_j)=
4(\bar{\vec u}_j)^2$, and we define $\bar r_j= (\bar{\vec u}_j)^2$ for all $j$. 
Taking into account this midpoint definition in the path integral (\ref{KlCy}) 
we obtain by performing the transformation $r\mapsto \vec u$ be means of
$A(\bar{\vec u}_j)$
\begin{eqnarray} & &\!\!\!\!\!\!
  K_l^C(r'',r';T)=\dfrac1{4r'r''}\int_{\bbbr}dy''
  \limN\bigg(\dfrac{2m}{\pi\i\hbar\epsilon}\bigg)^N
    \int_{\bbbr^2} (\bar{\vec u}_1)^2d\vec u_1\dots
    \int_{\bbbr^2} (\bar{\vec u}_{N-1})^2d\vec u_{N-1}\qquad
  \nonumber\\    & &\!\!\!\!\!\!\qquad\times
    \exp\Bigg[\ih\sum_{j=1}^N\Bigg(\dfrac{2m}{\epsilon}
    (\bar{\vec u}_j)^2(\Delta\vec u_j)^2
    -\dfrac{\epsilon\hbar^2l(l+1)}{2m (\bar{\vec u}_j)^2 (\bar{\vec u}_{j-1})^2}
    +\dfrac{\epsilon e^2}{ (\bar{\vec u}_j)^2}- \Delta V(\bar{\vec u}_j)
    \Bigg)\Bigg]\enspace.
\end{eqnarray}
Here, $ \Delta V(\bar{\vec u}_j)$ must be determined via \cite{LEEB}
\begin{equation}
  \Delta V(\bar{\vec u}_j)=\dfrac{\hbar^2}{8m}\sum_{\alpha,\beta,\gamma}
  \bigg(\dfrac{\partial}{\partial u_\alpha}
        \dfrac{\partial u_\beta}{\partial u_\gamma}\bigg) 
  \bigg(\dfrac{\partial}{\partial u_\beta}
        \dfrac{\partial u_\alpha}{\partial u_\gamma}\bigg)
  \bigg|_{\vec u=\bar{\vec u}_j}
  =\dfrac{\hbar^2}{8m (\bar{\vec u}_j)^4}\enspace.
\end{equation}
I perform a time-substitution on midpoints according to
$s_j-s_{j-1}=\Delta s_j=\epsilon/4(\bar{\vec u}_j)^2$ and obtain
\begin{equation}
  K^C_l(r'',r';T)=\int_{\bbbr}\dfrac{dE}{2\pi\i}\e^{-\i ET/\hbar}
  G^C_l(r'','r;E)\enspace,
\end{equation}
with the energy-dependent Green functions $ G^C_l(r'','r;E)$ given by
\begin{eqnarray} & &
  G^C_l(r'','r;E)=\dfrac{1}{\sqrt{r'{r''}^3}}\int_{\bbbr}dy''
  \ih\int_0^\infty ds''\,\e^{\i e^2s''/\hbar}
  \nonumber\\    & &\qquad\times
  \pathints{\vec u}\exp\Bigg[\ih\ints\Bigg(\dfrac{m}{2}\dotv{u}
  +E|\vec u|^2-\hbar^2\dfrac{(2l+1)^2}{2m|\vec u|^2}\Bigg)ds\Bigg]\enspace.  
\label{GlCu}
\end{eqnarray}
The path integral (\ref{GlCu}) is the path integral of a radial harmonic
oscillator in $\bbbr^2$. We introduce polar coordinates
\begin{equation}
\begin{array}{c}
u_1=v\cos\dfrac\alpha2\enspace,\\[3mm]
u_2=v\sin\dfrac\alpha2\enspace,\end{array}\qquad
0\leq\alpha<4\pi\enspace,\ v=|\vec u|\enspace, 
\end{equation}
use the expansion of plane waves into circular waves in terms of modified
Bessel functions $I_\nu$
\begin{equation}
 \e^{z\cos\psi}=\sum_{\nu\in\bbbz}\e^{\i\nu\psi}I_\nu(z)\enspace,
\end{equation}
and the asymptotic expansion \cite{FLMa,INOa,INOc}
\begin{equation}
  I_\nu\bigg(\dfrac{z}\epsilon\bigg)
  \mathrel{\mathop\simeq^{(\epsilon\to0)}}
  \sqrt{\epsilon\over2\pi z}\,\exp\bigg(\dfrac{z}\epsilon
       -\dfrac{\nu^2-1/4}{2z}\bigg)\enspace,
\end{equation}
and obtain after an $\alpha$-angular integration over $4\pi$ $\Big($we set
$\omega=\sqrt{-E/2m}\,\Big)$
\begin{eqnarray} & & \!\!\!\!\!\! \!\!\!\!\!\!
  G_l^C(r'',r';E)=\dfrac{2}{\sqrt{r'r''}}
  \int_0^\infty\dfrac{\e^{\i e^2s''/\hbar}s''}{\sqrt{v'v''}}  
  \nonumber\\    & & \!\!\!\!\!\! \!\!\!\!\!\!\qquad\times
  \pathints{v}\exp\Bigg[\ih\ints\Bigg(\dfrac{m}2\big(\dot v^2-\omega^2v^2\big)
      -\hbar^2\dfrac{(2l+1)^2-1/4}{2mv^2}\Bigg)ds\Bigg]
  \nonumber\\    & & \!\!\!\!\!\! \!\!\!\!\!\!
  = \int_0^\infty\dfrac{\e^{\i e^2s''/\hbar}s''}{\sqrt{v'v''}}  
  \dfrac{m\omega\sqrt{v'v''}}{\i\hbar\sin\omega s''}
  \exp\bigg[\dfrac{\i m\omega}{2\hbar}({v'}^2+{v''}^2)\cot\omega s''\bigg] 
  I_{2l+1}\bigg(\dfrac{m\omega v'v''}{\i\hbar\sin\omega s''}\bigg)
  \nonumber\\    & & \!\!\!\!\!\! \!\!\!\!\!\!
  =\dfrac1\hbar\dfrac1{r'r''}\sqrt{-\dfrac{m}{2E}}
  \dfrac{\Gamma(1+l-\kappa)}{(2l+1)!}
  W_{\kappa,l+\half}\bigg(\sqrt{-8mE}\,\dfrac{r_>}\hbar\bigg)
  M_{\kappa,l+\half}\bigg(\sqrt{-8mE}\,\dfrac{r_<}\hbar\bigg)\enspace.
\label{GlC}
\end{eqnarray}
Here, use has been made of the well-known path integral identity for the
radial harmonic oscillator \cite{PI}, 
\begin{eqnarray}       & &
  \pathint{r}\exp\left\{\ih\intt\Bigg[{m\over2}\Big(\dot r^2-\omega^2r^2
    \Big)-\hbar^2{\lambda^2-1/4\over2mr^2}\Bigg]dt\right\}
         \nonumber\\   & &
  =\sqrt{r'r''}{m\omega\over\i\hbar\sin\omega T}
  \exp\bigg[-{m\omega\over2\i\hbar}({r'}^2+{r''}^2)\cot\omega T\bigg]
  I_\lambda\bigg({m\omega r'r''\over\i\hbar\sin\omega T}\bigg)\enspace,
\end{eqnarray}
the integral representation \cite{GRA}
\begin{eqnarray} & &
  M_{\kappa\mu}(a) W_{\kappa\mu}(b)
  =\dfrac{\sqrt{ab}}{\Gamma)\mu-\kappa+\bhalf)}\int_0^\infty ds
  \bigg(\coth\dfrac{s}2\bigg)^{2\kappa} \qquad\qquad
  \nonumber\\    & &\qquad\qquad\times
  \exp\bigg(-\half(a+b)\cosh s\bigg)I_{2\mu}(\sqrt{ab}\,\sinh s)\enspace,
\end{eqnarray}
and I have used the usual abbreviation $\kappa=(e^2/\hbar)\sqrt{-m/2E}$.
$ r_\gtrless$ denotes the greater/lesser of $r',r''$.
From the poles of the Green function $G^C_l(E)$ we immediately derive the
Coulomb bound state spectrum
\begin{equation}
  E_n^C=-\dfrac{me^4}{2\hbar^2(n+l+1)^2}\enspace,\qquad n\in\bbbn_0\enspace.
\end{equation}  
The bound state wave-functions are obtained in the usual way by considering
the residua in (\ref{GlC}), and the scattering states are obtained by 
an analysis of the cuts of $G^C_l(E)$.

I have therefore obtained the radial Coulomb Green function by an
alternative method by using a singular version of the Levi-Civita 
transformation. The usual Levi-Civita transformation \cite{DKb,INOa} can be 
used in the evaluation of the two-dimensional Coulomb path integral.
Actually, the Levi-Civita transformation maps $\bbbr^2\mapsto\bbbr^2$:
$x=\xi^2-\eta^2,y=2\xi\eta$, and the coordinates $\xi,\eta$ correspond to
parabolic coordinates in $\bbbr^2$ in which the Coulomb problem is also
separable, e.g.~\cite{GROPOa} for an extensive discussion. For the singular
Kustaanheimo-Stiefel transformation in the case of the three-dimensional
Coulomb potential, the resemblance to parabolic coordinates can also
be recognised, and in the present case as well; however, in both cases we must
integrate over an additional variable. In all cases we encounter a path 
integral for a (radial) harmonic oscillator. Whereas in the three-dimensional
Coulomb potential the non-uniqueness of the transformation appears quite
naturally and leads to an isotropic harmonic oscillator in $\bbbr^4$, the
transformation (\ref{Au}) looks somewhat circumstantial, because the
one-dimensional transformation $r=u^2$ $(u>0)$ also does the job in a much
simpler way \cite{INOK}. The virtue of (\ref{Au}) lies in the fact that it 
demonstrates once again the importance of the Kustaanheimo-Stiefel 
transformation, it points out that singular transformations also work in the 
path integral, and that it emphasizes a singular transformation may be in 
other cases the only means which work.

\section*{Acknowledegement}
I would like to thank the members of the Joint Institute for Nuclear Research,
Dubna, in particular G.Pogosyan, for the warm hospitality during my stay in 
Dubna. I also thank F.Steiner for useful discussions on the Kepler-Coulomb
problem.
\newline
Financial support from the Heisenberg--Landau program is greatfully 
acknowledged.



\end{document}